\documentclass{PoS}

\usepackage{amsmath,amssymb,bm}

\usepackage{cite}

\title{Partonic quasi-distributions of the pion \\ in chiral quark models}

\ShortTitle{Pion quasi-distributions}

\author{\speaker{Wojciech Broniowski}\thanks{Supported by the Polish National Science Center grant 2015/19/B/ST2/00937.}\\
       The H. Niewodnicza\'nski Institute of Nuclear Physics, Polish Academy of Sciences, \\ 31-342 Cracow, Poland, and \\
       Institute of Physics, Jan Kochanowski University, 25-406 Kielce, Poland\\
       E-mail: \email{Wojciech.Broniowski@ifj.edu.pl}}

\author{Enrique Ruiz Arriola\thanks{Supported  by the Spanish Mineco grant FIS2014-59386-P, and by the
Junta de Andaluc\'{\i}a grant FQM225-05.} \\
        Departamento de F\'isica At\'omica, Molecular y Nuclear and Instituto Carlos I \\ de
  Fisica Te\'orica y Computacional,  Universidad de Granada, E-18071  Granada, Spain \\
        E-mail: \email{earriola@ugr.es}}

\abstract{The evaluation of partonic distributions presents a challenge for QCD, and in particular for its Euclidean lattice realization. 
Recently, objects called quasi-distributions (which become standard distributions in a limit of the longitudinal momentum of 
the target hadron going to infinity) have been proposed. We present a non-perturbative, dynamical 
evaluation of the quark quasi-distribution amplitude (QDA) of the pion in 
the framework of chiral quark models (the Nambu--Jona-Lasinio model and the spectral quark model). We arrive at simple but 
nontrivial analytic expressions, where the dependence on the longitudinal momentum, the momentum fraction, or the 
transverse-momentum (for the unintegrated objects) can be explicitly assessed. For the parton distribution amplitude (PDA), 
we carry out the necessary QCD evolution from the constituent quark model scale to higher scales accessible on the 
lattice, and compare favorably to the LaMET data.}

\FullConference{XVII International Conference on Hadron Spectroscopy and Structure - Hadron2017\\
		25-29 September, 2017\\ University of Salamanca, Salamanca, Spain}

\begin{document}

This talk presents a follow up of~\cite{Broniowski:2017wbr}, where more details of our approach 
can be found. Techniques used in this work
were also recently reported in~\cite{Broniowski:2017gfp}. The relevant lattice studies of the quasi-distributions in the pion 
were published in~\cite{Zhang:2017bzy}. 
The cousin problem of the Euclidean lattice extraction of the partonic quasi-distributions~\cite{Ji:2013dva} in the nucleon
attracted a lot of attention from the simulation side~\cite{Alexandrou:2015rja,Alexandrou:2016jqi,Alexandrou:2017qpu,Orginos:2017kos} 
as well as in theoretical developments~\cite{Xiong:2013bka,Ji:2014gla,Ma:2014jla,Ji:2015jwa,Ji:2015qla,Radyushkin:2016hsy,Monahan:2016bvm,Chen:2016fxx,%
Ji:2017rah,Chen:2017mzz,Carlson:2017gpk,Briceno:2017cpo,Rossi:2017muf,Stewart:2017tvs,Chen:2017lnm,Wang:2017qyg,Hobbs:2017xtq}. A 
community white paper has also just 
appeared~\cite{Lin:2017snn}. For brevity, we present here the distribution amplitudes only, however, the results 
are analogous for the distribution functions~\cite{Broniowski:2017wbr}.

The definition of the pion quasi-distribution amplitude (QDA)~\cite{Ji:2013dva} is given by the matrix elements of the bilocal quark operators,
\begin{eqnarray}
\tilde \phi(y,P_3) = \frac{i}{f} \int \frac{dz_3}{2\pi} e^{-i(y-1) z_3 P_3} 
 \left . \langle \pi(P) | \overline{\psi}(0) \gamma^3 \gamma_5 U(0,z) \psi(z) | 0 \rangle \right |_{z_0=0,\bm{z}_T=0}.
\end{eqnarray}
Similarly, the pion light-cone wave function (LCWF) is
\begin{eqnarray}
\Psi(x,\bm{k}_T) = \frac{i}{f} \int \frac{dz_-}{2\pi} e^{i(x-1) z_- P_+}  \int \frac{d^2z_T}{(2\pi)^2} e^{-i \bm{z}_T \cdot \bm{k}_T}
\left . \langle \pi(P) | \overline{\psi}(0) \gamma^+ \gamma_5 U(0,z) \psi(z) | 0 \rangle \right |_{z_+=0}.
\end{eqnarray}
Above, $z$ denotes the spatial separation of the quark operators (isospin indices are suppressed for brevity), $P^\mu$ is the four-momentum of the pion, 
$y \in (-\infty,\infty)$ is the  fraction of $P_z$, and $x \in [0,1]$ is the  fraction of the light-cone momentum $P_+$  
carried by the valence quark. The gauge link operator 
$U(0,z)$ is neglected in chiral quark models. 

Following the Lorentz covariance of the matrix elements of the quark bilinears, Radyushkin~\cite{Radyushkin:2017cyf} 
derived an important relation
\begin{eqnarray}
\tilde \phi(y,P_3) = \int_{-\infty}^{\infty} \!\!\!\! dk_1 \int_0^1 \!\!\! dx \, P_3  \Psi(x,k_1,(x-y) P_3). 
\end{eqnarray}
Thus, QDA can be obtained from LCWF via a straightforward 
double integration. From rotational symmetry, $\Psi(x,\bm{k}_T)=\Psi(x,k_T^2)$.
A similar methodology, also based entirely on the Lorentz covariance, was used in~\cite{Miller:2009fc,Broniowski:2009dt,Arriola:2010up,Miller:2010nz} 
to derive the {\em transversity relations} between the light-cone and equal-time
wave functions (Bethe-Salpeter amplitudes) of the pion.

\begin{table} 
\caption{Analytic formulas for various objects describing the valence quark distribution in the pion, evaluated in the NJL and SQM chiral quark models 
at the quark model scale and in the chiral limit, $m_\pi=0$. \label{tab:form}}
\begin{center}
\begin{tabular}{l|l|l} \hline
Name   & & \\
Symbol & \hspace{2.5cm} NJL & \hspace{2.1cm} SQM \\ \hline 
DA & & \\
$\phi(x)$ &  $\theta[x(1-x)] $ & $ \theta[x(1-x)]$ \\ \hline
QDA & & \\
 $\tilde \phi(y,P_3)$ & $\displaystyle \frac{N_c M^2}{4\pi^2 f^2} \left . \! {\rm sgn}(y) \ln  \frac{P_3 |y|\!+\!\sqrt{M^2\!+\!P_3^2 y^2}}{M} \right|_{\rm reg}$
                                                    & $\displaystyle \frac{1}{\pi} \left [  \frac{2 m_\rho {P_3} y}{m_\rho^2\!+\!4 {P_3}^2 y^2}
                                                                                                            \! +\!{\rm arctg} \left(\frac{2 {P_3} y}{m_\rho}\right) \right ]$\\
                             & $~~~~~~ + (y\leftrightarrow 1-y)$   & $~~~~~~ + (y\leftrightarrow 1-y)$ \\ \hline
LCWF & & \\              
 $\Psi(x,k_\perp)$ & $\displaystyle \frac{N_c M^2}{4\pi^2 f^2}  \left . \frac{1}{k_T^2+M^2} \right |_{\rm reg} \theta[x(1-x)]$  
                                                     & $\displaystyle  \frac{6 m_\rho^3}{\pi  \left(4 k_\perp^2+m_\rho^2\right)^{5/2}} \theta[x(1-x)]$ \\ \hline
pseudo-DA & & \\
${\cal P}(x,|\bm{z}|)$ & $\displaystyle \frac{N_c M^2}{4\pi^3 f^2} K_0(M |\bm{z}|) \bigg |_{\rm reg}  \theta[x(1-x)]$
                                                    &  $\displaystyle \frac{1}{2} e^{-\frac{ m_\rho |\bm{z}|}{2}} \left(m_\rho |\bm{z}|+ 2 \right) \theta[x(1-x)]$ \\ \hline
IDA & & \\
${\cal M}(\nu, |\bm{z}|)$ & $\displaystyle \frac{N_c M^2}{2\pi^3 f^2} \frac{\sin \left(\frac{\nu }{2}\right)}{\nu } K_0(M |\bm{z}|) \bigg |_{\rm reg}$ 
                                                    & $\displaystyle  \frac{\sin \left(\frac{\nu }{2}\right) }{\nu} e^{-\frac{ m_\rho |\bm{z}|}{2}} \left(m_\rho |\bm{z}|+ 2 \right)$ \\ \hline
VDA & & \\
$\Phi(x,\mu)$ & $\displaystyle \frac{N_c M^2}{4\pi^2 f^2} \mu \left . e^{-\mu M^2}  \right |_{\rm reg}  \theta[x(1-x)]$ 
                                                    & $\displaystyle \frac{\mu^{5/2}  m_\rho^3 e^{-\frac{1}{4}  \mu  m_\rho^2}}{4 \sqrt{\pi }} \theta[x(1-x)]$ \\ \hline
\end{tabular}
\end{center}
\end{table}

The standard parton distribution amplitude (PDA) of the pion is obtained as the limit~\cite{Ji:2013dva}
$\phi(x=y) = \lim_{P_3 \to \infty} \tilde \phi(y,P_3)$, where the support $x \in [0,1]$ is regained.
Other pertinent quantities are related to the above definitions via Fourier transforms. 
The {\em pseudo-distribution amplitude} (pseudo-DA)~\cite{Radyushkin:2017cyf} is given 
by the Fourier transform of the LCWF and noticing that from rotation invariance in depends in general on $|\bm{z}|$:
\begin{eqnarray}
{\cal P}(x,|\bm{z}|)=\int d^2 k_T   e^{i \bm{z}_T \cdot \bm{k}_T} \Psi(x,\bm{k}_T) ,
\end{eqnarray}
where $|\bm{z}|=z_T$. In 
particular, in the frame $P=(E,0,0,P_3)$ we may chose $z=(0,0,0,z_3)$, where $x$ is the Fourier conjugate variable of $P\cdot z=-P_3 z_3$. 
Next, the {\em Ioffe-time distribution amplitude} (IDA)~\cite{Braun:1994jq,Radyushkin:2017cyf} is simply related to the pseudo-DA:
\begin{eqnarray}
{\cal M}(\nu,|\bm{z}|) = \int_{0}^1 \! dx \, e^{i (x-\frac{1}{2}) \nu} {\cal P}(x,|\bm{z}|) \label{eq:Ioffe}
\end{eqnarray}
(we shift $x$ by ${\frac{1}{2}}$ to get real expressions).
Finally, the {\em virtuality distribution amplitude} (VDA)~\cite{Radyushkin:2014vla} is defined via (for simplicity of the resulting expressions 
we take a real Laplace transform)
\begin{eqnarray}
\Psi(x,k_\perp) = \frac{1}{\pi} \int_0^\infty \frac{d\mu}{\mu} e^{-k_\perp^2 \mu} \Phi(x,\mu).
\label{eq:VDmu}
\end{eqnarray}

\begin{figure}[tb]
\begin{center}
\includegraphics[angle=0,width=0.45 \textwidth]{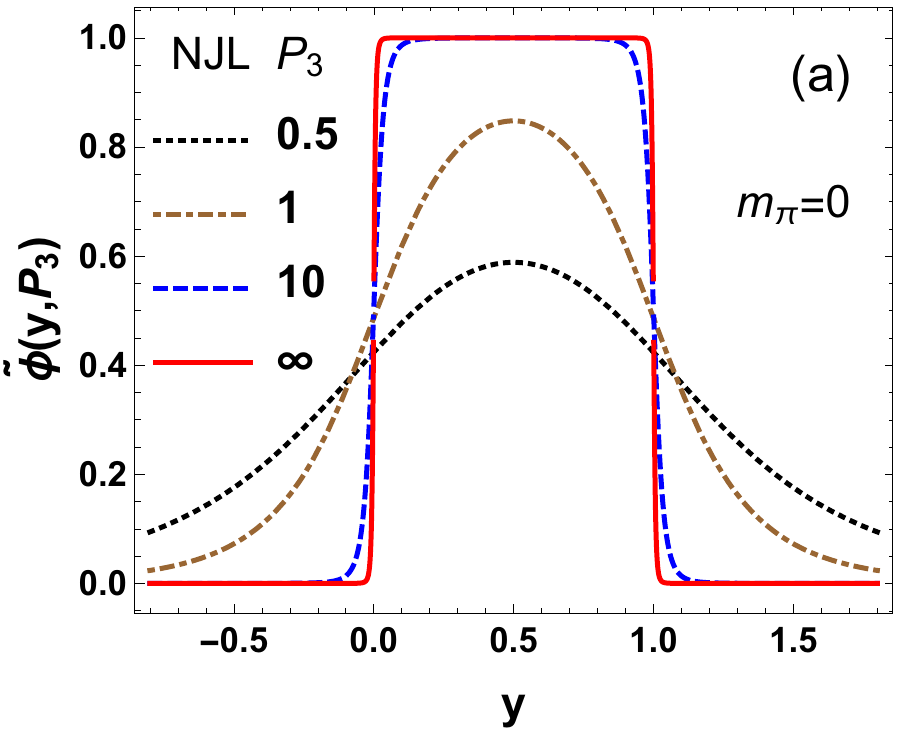} \hspace{.05\textwidth}  \includegraphics[angle=0,width=0.45 \textwidth]{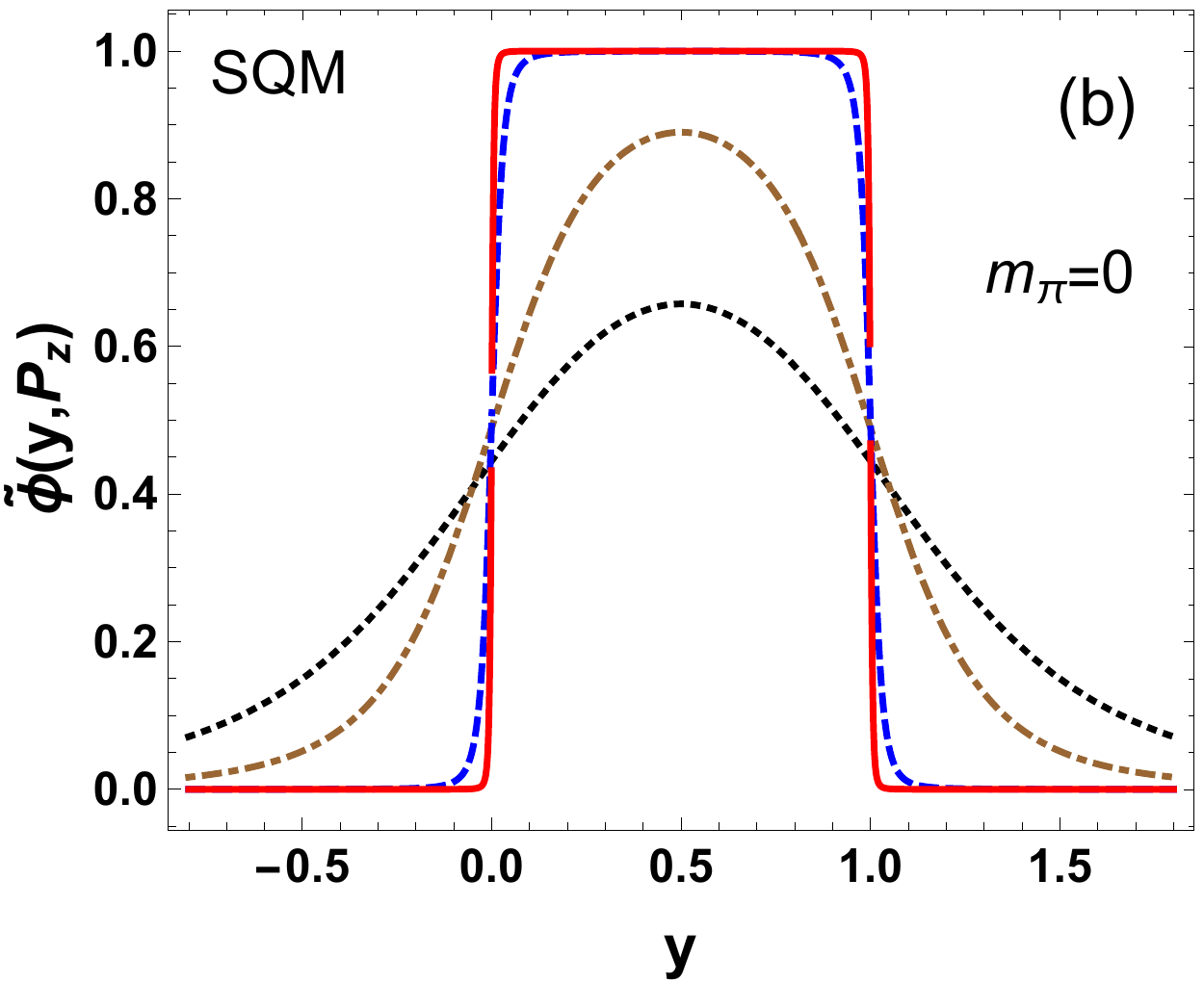}
\end{center}
\vspace{-6mm}
\caption{Valence quark quasi-distribution amplitude (QDA) of the pion in (a)~NJL and (b)~SQM models, obtained at the constituent quark scale $\mu_0$ with $m_\pi=0$ at 
various values of the longitudinal momentum $P_3$, plotted as a function of the longitudinal momentum fraction $y$. \label{fig:pz}} 
\end{figure} 

In chiral quark models, the evaluation of the valence quark distributions in the leading-$N_c$ order amounts to computing 
a one-quark-loop integral (for a review of techniques see, e.g., \cite{Broniowski:2007si}).
For the pion LCWF we get~\cite{RuizArriola:2002bp}
\begin{eqnarray}
\Psi(x,k_T^2) =   \frac{N_c M^2}{4\pi^2 f^2} \frac{\theta[x(1-x)]}{k_T^2+M^2 - m_\pi^2 x(1-x)} \bigg |_{\rm reg}. \label{eq:F}
\end{eqnarray}

The expression in Eq.~(\ref{eq:F}) must be properly regularized (i.e.,
in a way conserving the proper symmetries). For the NJL model we use
the Pauli-Villars regularization (see~\cite{Schuren:1991sc}).
The value of the cut-off is chosen in such a way that at a given value of the constituent quark mass $M$ (we use \mbox{$M=300$~MeV}) 
a proper value for the pion decay constant $f$ follows. 
In SQM, the regulator is imposed via a spectral integration over the quark mass along a suitably-chosen 
complex contour~\cite{RuizArriola:2001rr,RuizArriola:2003bs}. The prescription reproduces the 
phenomenological monopole shape
of the pion electromagnetic form factor, $F(t)=1/(1-t/m_\rho^2)$, where $m_\rho$ is the mass of the $\rho$ meson.

We note from Eq.~(\ref{eq:F}) that the longitudinal-transverse factorization 
at the quark model scale \mbox{$\mu_0\simeq 320$~MeV}~\cite{Broniowski:2007si}, i.e., the factorization between the $x$ and $k_T$ variables, holds only 
in the strict chiral limit.

\begin{figure}[tb]
\begin{center}
\includegraphics[angle=0,width=0.45\textwidth]{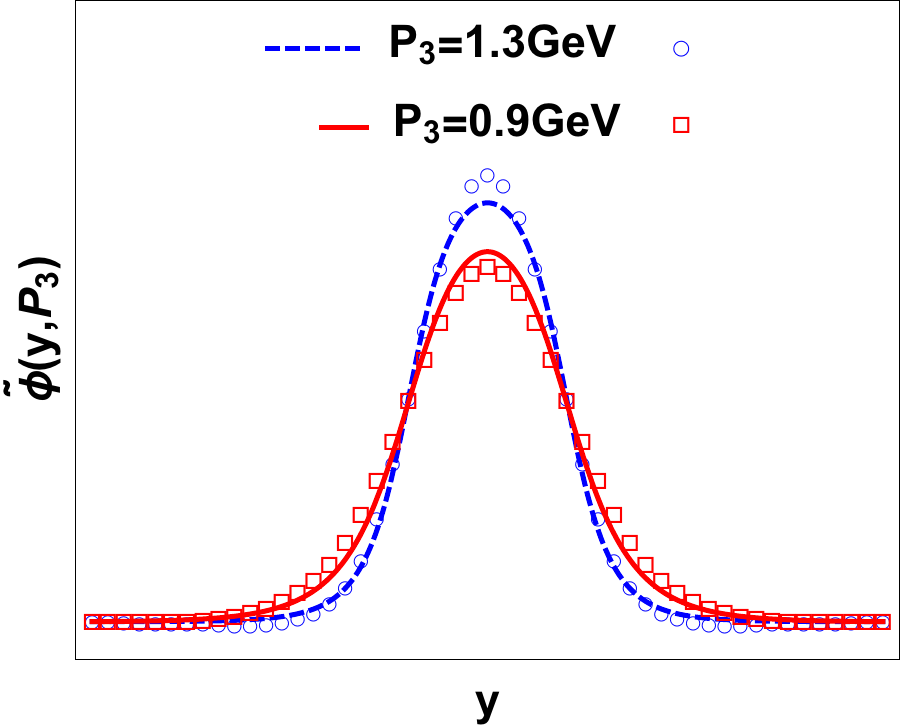} \hspace{.05\textwidth}  \includegraphics[angle=0,width=0.45\textwidth]{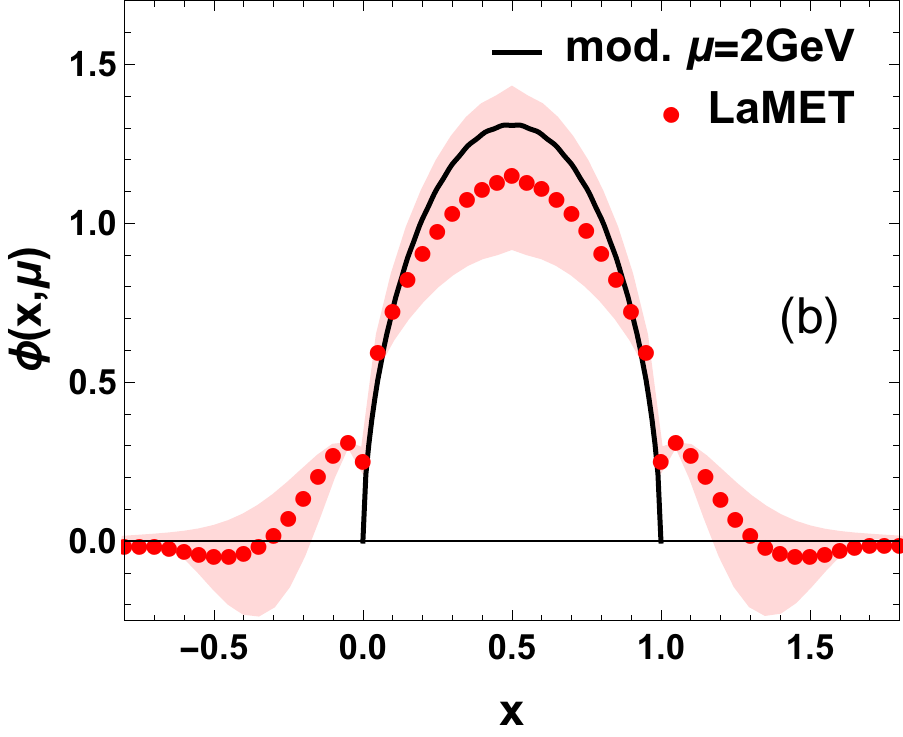}
\end{center}
\vspace{-6mm}
\caption{(a) Quark quasi-distribution amplitude (QDA) of the pion in the NJL model at the constituent quark scale $\mu_0$,
plotted as functions of the longitudinal momentum fraction $y$, evaluated for $P_3=0.9$ and $1.3$~GeV, and
compared to the lattice data at $\mu=2$~GeV from the LaMET Collaboration at $m_\pi=310$~MeV~\cite{Zhang:2017bzy}. 
(b)~The distribution amplitude (PDA) of the pion, obtained from the NJL  model and evolved to  
$\mu=2$~GeV, compared to the extraction from the LaMET data~\cite{Zhang:2017bzy}. \label{fig:DA}}
\end{figure} 

The results for the NJL and SQM models are analytic, but they are particularly simple in the chiral limit of $m_\pi=0$. 
They are collected in Table~\ref{tab:form}. The QDAs in the chiral limit at various values of $P_3$ are plotted in Fig.~\ref{fig:pz}.
We note almost identical results from both models.  Comparison to the LaMET data~\cite{Zhang:2017bzy}, very favorable taking into account the 
simplicity of the model, is made in Fig.~\ref{fig:DA}.

\bibliography{QDF-TMD}

\end{document}